\documentstyle[prd,aps,epsf,floats]{revtex}
\draft

\begin{document}
\title{Ultra-High Energy Cosmic Ray Propagation in the Local Supercluster}
\author{G\"unter Sigl}
\address{Department of Astronomy \& Astrophysics, The
University of Chicago, Chicago, IL 60637-1433}
\author{Martin Lemoine}
\address{DARC, UPR-176, CNRS,
Observatoire de Paris, 92195 Meudon C\'edex, France}
\author{Peter Biermann}
\address{Max-Planck Institute for Radioastronomy,
D-53010 Bonn, Germany}

\date{\today}
\maketitle

\begin{abstract}
We present detailed numerical simulations and analytical
approximations of the propagation of
nucleons above $10^{19}\,$eV in the Local Supercluster,
assuming that the ambient magnetic field is turbulent, and its strength 
$0.01\,\mu{\rm G}\,\lesssim B_{\rm rms}\lesssim 1\,\mu{\rm G}$. 
In such strong magnetic fields, protons in the low energy part of
the spectrum, 
$10^{19}\,{\rm eV}\,\lesssim E\lesssim E_{\rm C}$ diffuse, while the
higher energy particles, with $E\gtrsim E_{\rm C}$ propagate along
nearly straight lines. The magnitude of the transition energy $E_{\rm C}$
depends mainly on the strength of the magnetic field, the coherence
length, and the distance to the source; for
$B_{\rm rms}\simeq0.1\,\mu{\rm G}$, a largest eddy of length
$\sim10\,{\rm Mpc}$, and a distance to the source $\sim10\,{\rm Mpc}$,
$E_{\rm C}\simeq 100\,{\rm EeV}$.
Our numerical treatment substantially improves
on previous analytical approximations, as it allows to treat
carefully the transition between the two propagation regimes, as well as
the effects due to inhomogeneities expected on scales of a few
Mpc. We show that a
turbulent magnetic field $B_{\rm rms}\sim 0.1\,\mu{\rm G}$, close to
equipartition, would allow to reproduce exactly the observed spectrum
of ultra high energy cosmic rays, up to the highest energy observed,
for a distance to the source $d\lesssim 10\,{\rm Mpc}$, for the
geometry of the Local Supercluster, {\it i.e.} a sheet of thickness
$\simeq10\,{\rm Mpc}$. Diffusion, in this case, allows to
reproduce the high flux beyond the Greisen Zatsepin Kuzmin cut-off,
with a soft injection spectrum $j(E)\propto E^{-2.4}$. Moreover, the
large deflection angles at the highest energies observed, typically
$\sim10^\circ$ for the above values, would explain
why no close-by astrophysical counterpart could be associated with
these events.



\end{abstract}


\section{Introduction}

The structure and strength of extragalactic magnetic
fields are very poorly known at present. These fields 
likely consist of two components: A large-scale component
that permeates the whole universe and constitutes a primordial
relic of physical processes in the early universe, and a more
localized component which is associated with the structure
of galaxies, galaxy clusters, and superclusters. For the
primordial field only upper limits are known, most notably from
the limits on the Faraday rotation of radio waves from distant
powerful radio sources~\cite{Kronberg,Vallee}. For a field of r.m.s.
strength $B_{\rm rms}$ and coherence length $\lambda_c$, this
constraint is $B_{\rm rms}\lambda_c^{1/2}\lesssim10^{-9}\,{\rm G}\,
{\rm Mpc}^{1/2}$. For the component of the field that is uniform on 
the Hubble scale, this constraint reads
$B_{\rm c}\lesssim 10^{-11}\,{\rm G}$.
Theoretical scenarios for the origin of
this primordial component, {\it e.g.}, as relic fields produced by
the electric currents present during the electroweak or the
QCD phase transition~\cite{SJO}, or by the Biermann battery
effect~\cite{biermann}, predict much smaller field strengths,
typically of the order of $10^{-20}\,$G. In contrast, positive
detections from measurements of
Faraday rotation and synchrotron emission exist for the
localized component~\cite{Kronberg,Vallee}.
Estimates range from $10^{-9}\,$G
up to $10^{-7}\,$G over scales of Mpc, and in some cases
up to $10^{-6}\,$G, {\it e.g.}, in the core of galaxy
clusters~\cite{Kronberg}. Notably, a broad $\sim50^\circ$ region
centered on the Local Virgo Supercluster seems to be endowed with
a magnetic field with a uniform component of strength
$\sim1.5\times10^{-6}\,{\rm G}$~\cite{Vallee90},
and a random component of strength
$\sim10^{-6}\,{\rm G}$~\cite{Thomson82}. The coherence
length of random components is usually taken to be a cell size of
$\sim20\,{\rm kpc}$\cite{Vallee87}.

The origin of this component
is unclear, the most common explanations being the
action of a dynamo that exponentially amplifies seed fields
of the order of $10^{-20}\,$G in the flows of the large scale
galaxy structures, adiabatic compression of
strong primordial fields in the range $10^{-12}-10^{-9}\,$G,
or ejection from radiogalaxies~\cite{ensslin} or normal
galaxies~\cite{ryu}.

After several generations of experiments, most notably the Haverah
Park~\cite{Haverah}, the Akeno Giant Air Shower
Array (AGASA)~\cite{Yoshida1,Hayashida1}, and the Fly's
Eye~\cite{Bird1} experiments,
the origin and the nature of ultra high energy cosmic rays
(UHE CRs), with energies $E\gtrsim10\,$EeV
($1\,{\rm EeV}=10^{18}\,$eV) constitute an unsolved
problem as well. Nucleons above a few tens of EeV are
limited in their range to at most $\simeq50\,$Mpc, due
to the production of pions on the cosmic microwave
background (CMB), known as the Greisen-Zatsepin-Kuzmin (GZK)
effect~\cite{GZK}. No compelling astrophysical candidate for the
source of the highest energy events could be found
within this distance scale~\cite{ES,HVSV,biermann1}. Other primaries
such as nuclei, $\gamma-$rays, or neutrinos, pose
similar problems~\cite{SSB}. Theoretical scenarios for
their origin include powerful radio galaxies~\cite{biermann2},
cosmological $\gamma-$ray bursts~\cite{Waxman}, and
topological defects left over from phase transitions
in the early universe that took place at temperatures
near Grand Unification Scales~\cite{BHS,Sigl}.

Magnetic fields above $\sim10^{-12}\,$G on Mpc scales
significantly influence the propagation
of charged cosmic rays up to the highest energies ever
observed, namely several $100\,$EeV, by deflecting and
delaying them relative to straight line propagation with
the speed of light. The absence of
a significant correlation of their arrival directions
with the Galactic plane~\cite{Stanev1} and the fact that
protons at these energies cannot be confined within the
Galactic magnetic field, suggest that
UHE CRs above $\sim1\,$EeV
most likely have their origin in extragalactic sources.

We recently developed a Monte Carlo code
for the propagation of nucleons in extragalactic magnetic fields.
We demonstrated that a combination of this code with a maximum
likelihood approach allows to extract from observed distributions
of arrival directions, times, and energies information on the
nature of the source mechanism as well as on the strength and
structure of large-scale cosmic magnetic fields,
without taking recourse to any specific scenario for the origin of
UHE CR~\cite{SL,S_OWL}.
This was discussed in a qualitative way in Refs.~\cite{LSOS,SSLCH}
and applied to the pairs of events recently reported by the
AGASA experiment~\cite{Hayashida1} in Ref.~\cite{SLO}.
Our studies were motivated by the fact that
future large scale experiments~\cite{proc}, such as the High
Resolution Fly's Eye~\cite{Bird4}, the Telescope Array~\cite{Teshima},
and most notably the Pierre Auger project~\cite{Cronin}, should 
allow to detect clusters of $\gtrsim20$, and possibly more, UHE 
CRs per source, if the clustering suggested by the AGASA
results~\cite{Hayashida1} is a real effect caused by discrete
sources. The recently proposed satellite observatory
concept for an Orbital Wide-angle Light collector (OWL)~\cite{OWL}
might even allow to detect clusters of hundreds of events by
watching the Earth's atmosphere from space.

In our studies we adopted a statistical description of the
extra-galactic magnetic field
as a Gaussian random field in Fourier space with zero mean
and a power spectrum with $\left\langle B^2(k)\right\rangle\propto
k^{n_B}$ for $k<2\pi/l_c$ and $\left\langle B^2(k)\right\rangle=0$
otherwise, where $l_c$ is the numerical cut-off scale. Our studies
were thus performed for a ``homogeneous''
magnetic field, and, furthermore, assuming near straight line
propagation, {\it i.e.} $B_{\rm rms}\lesssim10^{-9}\,{\rm G}$.
The subject of the
present work is to generalize this to the case of more complex
structures of cosmic magnetic fields, most notably the possibility
of a locally enhanced magnetic field associated with the Local
Supercluster. Indeed, Ryu, Kang \& Biermann~\cite{RKB98} have recently
argued that the upper limit on the strength of a homogeneous magnetic
field, obtained from Faraday rotation measures of distant quasars,
becomes $B\lesssim1\mu{\rm G}$ in filaments and walls
of large scale structure, under the very reasonable assumption
that the magnetic field scales with the matter density.
On the other hand, if the two paired events
observed by AGASA close to the Supergalactic plane are due to
discrete sources and not a statistical fluctuation, the field
strengths in this structure must be smaller than $\sim10^{-9}\,$G.
However, Biermann et al.~\cite{BKRR} have argued that due to the
accretion flow with
embedded magnetic fields towards the Supergalactic plane,
weak confinement is possible, in which case the AGASA pairs
could have originated in unrelated sources.
The existence of strong magnetic fields in the local
Supercluster is therefore an unsettled issue.

Particles above $\sim10\,$EeV, that propagate over
cosmological distances in magnetic fields as strong as 
$B\gtrsim{\rm few}\times10^{-8}\,$G, enter the diffusion regime in which
they turn around several times before reaching the observer.
However, above a
certain energy $E_{\rm C}$, whose value depends on the strength of the
magnetic field, on the coherence length, and on the distance to the source,
charged UHE nucleons do not
diffuse, and rather propagate in a near straight line. This scenario,
proposed by Wdowczyk \& Wolfendale~\cite{WW79}, has several attractive
features with respect to the observed UHE CR spectrum.
The transition between diffusion
and near straight line propagation could {\it a priori} explain why the
UHE CR spectrum appears flatter above $\sim100\,{\rm EeV}$ than below,
although this trend is not confirmed due to a lack of statistics at
high energy. Moreover, such strong magnetic fields would
result in large deflection angles, even at the highest energies
observed, and would thus explain why no astrophysical counterpart
could be associated with these events. Finally, the work of 
Ryu, Kang \& Biermann~\cite{RKB98} seems to support the idea that the
magnetic field could be much stronger within walls and filaments of
large scale structure, than in the voids. The spectra of UHE CRs
in such structures have to some extent been discussed for
the diffusive limit in Ref.~\cite{AC94}

Recently, the diffusion of UHE CRs in the local supercluster has
been reconsidered in Ref.~\cite{BO}. These authors combined the
diffusion approximation at low energies with the straight
line propagation approximation at high energies, assuming
continuous energy losses and an infinitely thick supergalactic plane.
The main difference to Ref.~\cite{WW79} lies in a better
treatment of energy losses, a more careful treatment of the transition
regime between diffusion and rectilinear propagation, a more quantitative
appraisal of the diffusion coefficient, and a more detailed discussion
with respect to present data. One goal of the present work is to
overcome the approximations made in Refs.~\cite{WW79,BO}, {\it i.e},
to provide more accurate flux calculations at ``intermediate'' energies
where both the diffusion and the small
deflection approximations break down, and to examine the effect of an
inhomogeneous distribution of densities and magnetic fields in
the local Supercluster. This latter effect is
especially important as the distance to the nearest cluster (Virgo,
$d\sim20$Mpc) is already larger than the typical scale of these
inhomogeneities for which a few Mpc is a reasonable assumption.

We describe our approach in Section~2, present results of the
numerical simulations in Section~3, and conclude in Section~4.
Natural units are used throughout, i.e., $c=\hbar=1$.

\section{Numerical Approach}
A detailed account of our approach was given in Refs.~\cite{SLO,SL}.
Here, we restrict ourselves to a brief description of the most
important aspects and the generalizations relevant for the present
study.

A specific scenario is characterized by the magnetic field
realization and the position of source and observer. For each
such scenario, we computed a few $10^3$ trajectories of nucleons
emitted from a given source in random directions within a cone
with opening angle of typically $90^\circ$ and with a flat injection
spectrum. This was done by solving the equations of motion in the
presence of the magnetic fields, taking into account pion
production losses by sampling the relevant differential cross
sections, as well as losses from pair production by protons on
the CMB as a continuous energy loss process. 
If the nucleon crossed the sphere of radius $d$
around the source on a cap with an opening angle of typically
a few degrees, centered on the observer, the
nucleon was considered as arriving to the observer. In this case its
injection energy, and the energy, time, and propagation direction
upon arrival, the latter defined relative to the line of sight
to the source, were recorded. The trajectory of the nucleon does not
stop here, however, as, in the diffusion regime, one needs to
calculate the number density of particles at a given point.
Therefore, in our calculations, a given nucleon propagates until 
either the propagation time exceeded $10^{10}\,$yr,
or the distance exceeded twice the distance $d$ between the source and
the observer, and, on its way, it can give multiple hits on the
``detector''.  Since particles crossing the cap
from either side were considered as arriving, the maximal deflection
angle is $180^\circ$.
The contribution of a particular particle to the flux per area
is weighted by the inverse of the absolute value of the cosine
of the arrival direction relative to the line of sight to the
source. If the linear dimension of the cap centered
at the observer is comparable or smaller than the typical gyro
radius of the particle and the coherence length of the magnetic
field, and the emission cone angle is larger than typical deflection
angles, this procedure ensures a realistic distribution of
arrival energies, times, and directions without too excessive
use of CPU time. By varying the cone angles within these limits,
we verified the independence of the results of the exact cone
angle values. Our procedure applies to arbitrary magnetic field
strengths and deflection angles.

The resulting distributions of UHE CRs can be folded with
the source emission timescale $T_{\rm S}$ in time and with the
injection spectrum in energy to obtain distributions for arbitrary
$T_{\rm S}$ and injection spectra.
Since we consider turbulent magnetic fields, defined by a Kolmogorov
spectrum, the spatial configuration of the magnetic field is highly
anisotropic, as seen from, either the source, or the observer. We thus
perform $\sim10$ runs for a given set of parameters, in order to
circumvent, and, at the same, examine, the effect of a particular
realization of the field on the spectrum. 

We approximate the structure of the local Supercluster as an
infinitely extended
two-dimensional sheet of half-thickness $d_s$ in the $x-y$ plane
at a position $z_s$. For the density profile $\rho(z)$ we
assumed a Gaussian of the form
\begin{equation}
  f(z)\equiv\rho(z)/\rho_0=\left\{\frac{1}{c}+
  \exp\left[-\left(\frac{z-z_s}{d_s}\right)^2\right]\right\}
  \,,\label{profile}
\end{equation}
where $c$ is the density contrast and $\rho_0$ is the density
at the center of the sheet. It is possible to include a further 
radial dependence of the density profile; however, for the sake of
simplicity, we will ignore such dependence in the following and
take the sheet to be uniform in the $x-y$ plane.

The random component of the magnetic field is then modeled
by using the Fourier transformed Gaussian field of strength
$B_{\rm rms}$ with the power law
spectrum described earlier, truncated below $k_{\rm min}=2\pi/L$,
with $L$ the largest turbulent eddy, and above
$k_{\rm max}=2\pi/l_c$, and multiplying it
with the profile $f(z)$. This does not strictly speaking
obey ${\bf \nabla\cdot B}=0$, but we do not expect this to
significantly influence the calculated histograms. For the
power law index we choose $n_B=-11/3$, corresponding to
a turbulent Kolomogov spectrum.

In addition, the continuity equation and Maxwell's equations
allow for a coherent field component ${\rm B}_c$ that is
parallel to the sheet and whose only dependence is on $z$
and is proportional to $f(z)$. We have verified in our Monte Carlo
simulations that a coherent component of strength considerably
smaller than the random component does not change results
significantly. For this reason we will neglect a coherent
component altogether in the rest of this paper.

Strictly speaking, due to the high conductivity of the
extragalactic medium, the accretion flow onto the sheet
causes an electric field ${\bf E}=-{\bf v}\times{\bf B}$,
where ${\bf v}$ is the velocity field. Because $v\ll1$,
we will, however, neglect this field in the present simulations.
Other interactions, such as with baryons, are also
negligible. We note, however, that solar modulation of
cosmic rays depends on this electric field, which thus
may be necessary to include in future studies.

In our simulations we usually consider one discrete source located at the
origin of the coordinate system, and the observer location
is characterized by its $z$ coordinate $z_o$, and its distance $d$
to the source; the source itself is characterized by a $z$ coordinate
$z_{\rm s}$.
In order to reduce the parameter space, we assume, in all
simulations, a sheet thickness $d_s=5\,$Mpc, a largest turbulent
eddy of size $L\sim2d_s$, and a magnetic field cut-off scale
$l_c=L/10$. The latter is limited by resolution but
does not  influence the result as long as $l_c\ll L$ because the
magnetic field power is concentrated at the largest scales. For
this reason, the effective coherence scale of magnetic fields
with a Kolomogorov spectrum is given by $\lambda_c\simeq L/(2\pi)$.
We further assume power law differential injection
spectra $Q(E)\propto E^{-\gamma}$, extending up to
$E=E_{\rm max}=10^4\,$EeV, but we note that
because of the falling fluxes the results and conclusions
are independent of maximal energies $E_{\rm max}$ for
$E_{\rm max}\gtrsim10^3\,$EeV.

\section{Results}

  Here, unless otherwise stated, we assume a stationary
situation, {\it i.e.} such that the emission timescale of the source
is much longer than the diffusion timescale (typically
$\sim10^8\,$yr). Sources with smaller emission time scales
tend to produce spectra that are peaked at specific energies,
and, therefore, individually provide much poorer fits to the
combined data above $\sim10\,$EeV.

\subsection{Diffusion of UHE CRs in an infinite medium}

  We first recall the effect of diffusion in an infinite medium on
charged UHE CRs. A similar discussion of this topic is given in
Ref.~\cite{BO}. Ignoring energy losses for a moment, there will
be a transition energy $E_{\rm C}$ such that particles with
$E\gg E_{\rm C}$ will propagate along nearly straight lines,
whereas particles with $E\ll E_{\rm C}$ will diffuse. This
transition occurs where the (energy-dependent) time delay $\tau_E$
with respect to rectilinear propagation with the speed of light
becomes comparable to the distance $d$. Since in the rectilinear
regime, $E\gtrsim E_{\rm C}$, the time delay is given by
\begin{equation}
  \tau_E\simeq\frac{2}{5}\frac{d^2L}{18r_g^2}\simeq6\times10^3
  \left(\frac{E}{100\,{\rm EeV}}\right)^{-2}
  \left(\frac{d}{10\,{\rm Mpc}}\right)^{2}
  \left(\frac{L}{10\,{\rm Mpc}}\right)
  \left(\frac{B_{\rm rms}}{10^{-9}\,{\rm G}}\right)^2\,{\rm yr}
  \,,\label{delayrect}
\end{equation}
assuming $n_B=-11/3$,
where $r_g=E/(eB_{\rm rms})$ is the gyro radius ($e$ is the
electron charge), this condition translates into
\begin{equation}
  E_{\rm C}\simeq150\,\left(\frac{d}{10\,{\rm Mpc}}\right)^{1/2}
  \,\left(\frac{B_{\rm rms}}{10^{-7}\,{\rm G}}\right)\,
  \left(\frac{L}{10\,{\rm Mpc}}\right)^{1/2}\,{\rm EeV}\,.
  \label{Ecrit}
\end{equation}
This estimate is confirmed by our numerical simulations within
a factor of about 2, for our parameters chosen.
For $E\gtrsim E_{\rm C}$, the observable differential spectrum
per solid angle (current density)
$j(E)$ is unmodified by the magnetic field with respect to the
injection spectrum and given by
\begin{equation}
  j(E)=\frac{Q(E)}{(4\pi)^2 d^2}\,.\label{fluxrect}
\end{equation}

The diffusive regime, $E\lesssim E_{\rm C}$, is usually described
by an energy-dependent diffusion coefficient $D(E)$, and the
expressions corresponding to Eqs.(\ref{delayrect}) and
(\ref{fluxrect}) read
\begin{equation}
  \tau_E\simeq\frac{d^2}{D(E)}\,,\label{delaydiff}
\end{equation}
and
\begin{equation}
  j(E)=\frac{Q(E)}{(4\pi)^2\,d\,D(E)}\,,\label{fluxdiff}
\end{equation}
respectively.

Another way of defining the transition energy $E_{\rm C}$ is to
equate Eqs.~(\ref{fluxrect}) and (\ref{fluxdiff}) which gives
$d=D(E_{\rm C})$. For the Bohm diffusion coefficient
$D_{\rm Bohm}(E)=r_g/3$, this results in
$E_{\rm C}\simeq3.3\times10^3(d/10\,{\rm Mpc})
(B_{\rm rms}/10^{-7}\,{\rm G})\,$EeV, about a factor 20 higher than
Eq.~(\ref{Ecrit}) for our parameters. These estimates suggest
that the transition regime between the analytically tractable
limiting cases of diffusion and rectilinear
propagation may stretch more than one order of magnitude in energy and
further emphasizes the need for detailed numerical simulations
in that regime.

For a diffusion coefficient that
scales as $D(E)\propto E^m$, one thus obtains
$j(E)\propto E^{-\gamma-m}$, in the absence of energy losses.
From the approximate analytical expression
\begin{equation}
  D(E)\simeq\frac{1}{3}\,r_g(E)\,
  \frac{B_{\rm rms}}{\int_{1/r_g(E)}^\infty
  \,dk\,k^2\left\langle B^2(k)\right\rangle}\,,\label{Banaly}
\end{equation}
where $B_{\rm rms}^2=\int_0^\infty\,dk\,k^2\left\langle B^2(k)
\right\rangle$,
one expects $m\sim1$ for $r_g\gtrsim L/(2\pi)$ from Bohm
diffusion, and $m\sim1/3$ for $r_g\lesssim L/(2\pi)$ in the case of the
Kolmogorov spectrum. As a consequence,
the correlation between the time delay $\tau_E$ and
$E$ should switch from, roughly, $\tau\propto E^{-1/3}$ at largest time
delays, hence smallest energies, as long as $r_g\lesssim L/(2\pi)$, to
$\tau_E\propto E^{-1}$ in the regime of Bohm diffusion,
$r_g\gtrsim L/(2\pi)$, and eventually to $\tau\propto E^{-2}$ at
the smallest time delays, hence largest energies, in the
near straight line propagation regime, $E\gtrsim E_{\rm C}$.
As we discuss below, our simulations indeed show all three
propagation regimes, up to the difference that the diffusing regime with
$D(E)\propto E^{1/3}$ takes place when $r_g<0.1L/(2\pi)$, and
$D(E)\propto E$ otherwise as long as diffusion holds.

\begin{figure}[ht]
\centering\leavevmode
\epsfxsize=5.0in
\epsfbox{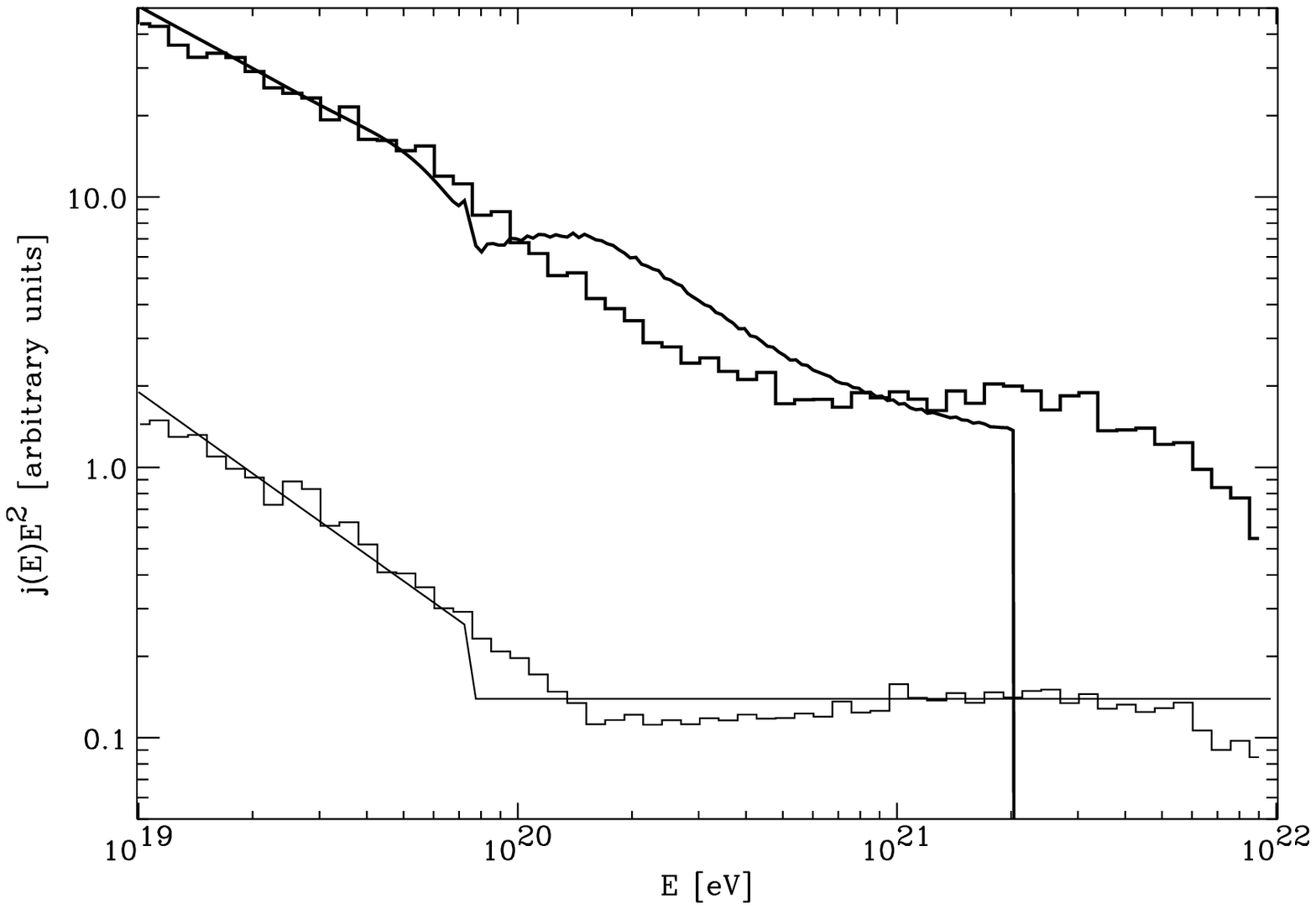}
\bigskip
\caption[...]{Spectrum of UHE CRs propagating in an infinite medium,
in a turbulent Kolmogorov magnetic field of strength
$B_{\rm rms}\simeq5\times10^{-8}\,{\rm G}$, for a distance to the
source $d=10\,{\rm Mpc}$, a largest turbulent eddy size
$\sim10\,{\rm Mpc}$, and a numerical cut-off scale
$l_c=L/10$. The injection spectrum is $Q(E)\propto E^{-2}$; the
spectrum shown is $E^2 j(E)$, with $j(E)$ the observable differential
flux, such that the deviation from a horizontal line shows the
modification of the injection spectrum.
The different cases shown are: no energy loss,
in thin solid histogram; energy losses, {\it i.e.} pair production and
stochastic pion production, in thick solid histogram (arbitrarily
shifted upwards from the no-energy loss case); analytical
approximations to each of the two spectra, obtained from solutions
of the equations discussed in the text, in light solid lines (the upper
solid line corresponds to the case where energy losses are included, as
for the Monte-Carlo results). All fluxes have been normalized to the
same total number of 'detected' particles. Analytical and numerical
simulations, with energy losses included, differ at high energy
$E>300\,$EeV due to the difference between continuous and stochastic
energy losses.}
\label{F1}
\end{figure}

The above picture is easily derived from analytical calculations.
However, such calculations cannot reproduce very well the
transition between the diffusive regime and the regime in which
particles propagate nearly linearly. In Figure~\ref{F1}, we show, as the thin
solid histogram, the spectrum obtained
for UHE CRs propagating in a $B_{\rm rms}\simeq5\times10^{-8}\,{\rm G}$
magnetic field, over a distance $d=10\,{\rm Mpc}$, in an infinite
medium, in the absence of energy losses, and for an injection spectrum
$\propto E^{-2}$. In the low energy region, 
$10\,{\rm EeV}\lesssim E \lesssim 70\,{\rm EeV}$, where Bohm
diffusion occurs, one recovers the slope predicted by Eq.~(\ref{fluxdiff}),
$j(E)\propto E^{-2}/D(E)\propto E^{-3}$. At high energy, 
$E\gtrsim 100\,{\rm EeV}$, where particles propagate in near straight
line, one recovers the injection spectrum, {\it i.e.}, the spectrum is
not modified during propagation. The transition between the two regimes
indeed occurs smoothly, as would be obtained from a simple
interpolation. Our simulations allow us to derive the magnitude of
the diffusion  by equating
the ratio of numerical fluxes in the diffusive and rectilinear
regimes with the ratio of Eqs.~(\ref{fluxdiff}) and (\ref{fluxrect}).
It is thus not set {\it a priori}; we obtain, for this latter: 
$D(E)\simeq3\times10^{32}\,{\rm cm}^2{\rm s}^{-1}\,(E/1\,{\rm EeV})
(B_{\rm rms}/1\,\mu{\rm G})^{-1}$, {\it i.e.} a factor $\simeq10$
higher than the Bohm diffusion coefficient. This is consistent
with the above mentioned fact that we observe Bohm diffusion for
$r_g>0.1L/(2\pi)$ and suggests that the magnetic field structure
in the numerical simulation leads to an effective gyro radius that
is about a factor 10 higher than the analytical expression
$r_g=E/(eB_{\rm rms})$.

We now turn to the effects caused by energy losses.
Their overall effect in the Monte Carlo
simulations is shown in Figure~\ref{F1}
as the thick solid histogram, for the same physical parameters as above.
The qualitative effect of
the introduction of pion production and pair production losses, for
$d\simeq10\,{\rm Mpc}$, is to change the slope to roughly 
$j(E)\propto E^{-2.6}$ ($\gamma=2$),
and to deplete the flux at high energies, in
favor of the flux at low energies, as pion production occurs mainly
above the GZK cut-off at $\sim50\,{\rm EeV}$.

Analytically, energy losses are often approximated as being
continuous, in which case the rate of change of the particle
energy is governed by the equation $dE/dt=b(E)$.
We note that this is not necessarily a good approximation
because the dominant energy loss process, namely pion
production, is purely stochastic and not continuous.
Our Monte Carlo simulations do not make this
approximation. 

An often used analytical expression for the
flux from a point source at distance $d$
in the rectilinear regime~\cite{BG} which generalizes
Eq.~(\ref{fluxrect}) is then given by (we neglect redshifting)
\begin{equation}
  j(E)=\frac{Q[E_i(E,d)]}{(4\pi)^2 d^2}\,\frac{dE_i(E,d)}{dE}
  \,,\label{fluxrectloss}
\end{equation}
where $E_i(E,r)$ satisfies the equation $\partial_r E_i(E,r)=-b[E_i(E,r)]$
with $E_i(E,d)=E$; $dE_i(E,d)/dE$ is the ratio of energy
intervals at emission and observation.

In the diffusive regime the problem can be
described by a partial differential equation for the local density
$n(E,{\bf r})=4\pi j(E,{\bf r})$ of UHE CRs of the following form
\begin{equation}
  \partial_t n(E,{\bf r})+\partial_E
  \left[b(E)n(E,{\bf r})\right]-
  \hbox{\boldmath$\nabla$}\left[D(E)
  \hbox{\boldmath$\nabla$}n(E,{\bf r})\right]=
  q(E,{\bf r})\,,\label{cel_d}
\end{equation}
where $q(E,{\bf r})$ is now the differential injection rate per
volume, $Q(E)=\int d^3{\bf r}\,q(E,{\bf r})$. Eq.~(\ref{cel_d})
also holds if $D(E)$ is space dependent (see next section).
In the homogeneous case, an analytical solution of Eq.~(\ref{cel_d})
is known~\cite{B91} which for a point source can be written as
\begin{equation}
j(E)=\frac{1}{4\pi}
\int_{0}^{+\infty}dE^\prime\,\frac{Q(E^\prime)}{|b(E)|
(4\pi\lambda^{2})^{3/2}}\,
\exp\left[-\frac{d^2}{4\lambda(E,E^\prime)^2}\right]\,,\label{Eq_diff_inf}
\end{equation}
where
\begin{equation}
\lambda(E_1,E_2)\equiv \left[\int_{E_2}^{E_1}dE^\prime\,
\frac{D(E^\prime)}{|b(E^\prime)|}\right]^{1/2}\, ,
\end{equation}
denotes an effective path length against losses
[$\lambda(E_1,E_2)\equiv0$ if $E_1>E_2$].

In the simplest approximation, Eq.~(\ref{fluxrectloss}) and
analytical solutions of Eq.~(\ref{cel_d}) are just used
for $E>E_{\rm C}$ and $E<E_{\rm C}$, respectively~\cite{WW79,BO}.
For the diffusive solution, the above integral should then be truncated
at $E_{\rm C}$.
A somewhat better analytical approximation can actually be
obtained by treating particles as propagating rectilinearly
as long as $E>E_{\rm C}$ {\it in flight}, but diffusively
otherwise. For a point source, this amounts to adding in ``secondary
sources'' in Eq.~(\ref{cel_d}), corresponding to the spatial distribution
of high energy particles, with $E>E_{\rm C}$ at injection, that have
propagated rectilinearly and lost energy down to $E_{\rm C}$.
The source term is thus replaced by:
\begin{equation}
  q(E,{\bf r})\longrightarrow q(E,{\bf r})\Theta(E_{\rm C}-E)+
  \int d^3{\bf r_i}
  \frac{q[E_i(E_{\rm C},|{\bf r}-{\bf r_i}|),{\bf r_i}]}
  {4\pi |{\bf r}-{\bf r_i}|^2}
  \frac{dE_i(E_{\rm C},|{\bf r}-{\bf r_i}|)}{dE}
  |b(E_{\rm C})|\delta(E-E_{\rm C})
  \label{subst}
\end{equation}
on the r.h.s. of Eq.~(\ref{cel_d}), where $\Theta(x)$ is the
Heaviside function,  ${\bf r_i}$ denotes the ``secondary'' source
location; for a point-like source, 
$q(E,{\bf r})=Q(E)\delta^3({\bf r}-{\bf r_s})$.
This  changes
the flux towards higher values if $E_{\rm C}$ is larger than
the GZK cutoff $\sim50\,$EeV.

We show the analytical predictions Eqs.~(\ref{fluxdiff})
and~(\ref{fluxrect}) (in the absence of energy losses),
and Eqs~(\ref{cel_d}), (\ref{subst}), and (\ref{fluxrectloss})
(including energy losses) as light and thick solid lines respectively,
in Fig.~\ref{F1},
for comparison with the Monte Carlo results. The continuous
energy loss rate $b(E)$ was taken from Refs.~\cite{BG,Blumenthal}. The
diffusion coefficient that enters the analytical solution of the
diffusion equation has been normalized to the value obtained from the
numerical solutions for a homogeneous medium in the absence of energy
losses. We note that the analytical solution, when energy losses are
included, does not reproduce correctly the transition between the two
different propagation regimes. As would be naively expected, the
energy losses around the GZK cut-off tend to smooth out the spectrum
in this transition region. Up to this inadequacy of the sudden
transition approximation, and the difference in energy losses (obvious
at high energy), there is good agreement between the
two methods, in this standard case.

\subsection{Diffusion of UHE CRs in the Local Supercluster}

  When diffusion occurs in a sheet of finite thickness, the above 
picture is significantly modified. For a better comprehension of
the different physical effects that enter the discussion,
we distinguish here two
different regimes: one in which the diffusive regime takes place only
for particles with energy less than the GZK threshold $\simeq50\,$EeV,
and the other in which particles with energies greater than the
GZK cut-off also diffuse. For a distance $d\simeq10\,{\rm Mpc}$, and
the above choice of parameters, these two regimes roughly correspond
respectively to $B_{\rm rms}\lesssim 5\times10^{-8}\,{\rm G}$, and 
$B_{\rm rms}\gtrsim 10^{-7}\,{\rm G}$. 

  We first discuss the case where diffusion occurs only below the GZK
cut-off. Since the observed UHE CR spectrum between $\sim5\,$EeV up
to the GZK cut-off can be fit with one power law, we select 
$B_{\rm rms}\simeq5\times10^{-8}\,{\rm G}$ as a fiducial value, in
which case particles up to the GZK cut-off diffuse, for a source
distance $d\sim10\,{\rm Mpc}$.
The main effect due to the finite thickness of the sheet is
that particles can escape freely at the boundary of the sheet, since
the magnetic field outside is too weak to turn the particles around. 
Therefore, the energy spectrum of UHE CRs is depleted in the energy
range where particles diffuse, whenever the source and/or the
observer are located near the boundary, or when the distance to the
source is much larger than the sheet thickness. It is important to
note that this effect occurs only for those particles that diffuse, as
the high energy particles that travel in near-straight line are not
affected by the presence of the boundary. 
We recall that diffusion of UHE CRs in the local Supercluster was
precisely suggested as a mechanism to enhance the local density of low
energy particles, with respect to that of high energy particles. We
thus see that the free escape of UHE CRs at the sheet boundary tends
to cancel this enhancement. 

\begin{figure}[ht]
\centering\leavevmode
\epsfxsize=5.0in
\epsfbox{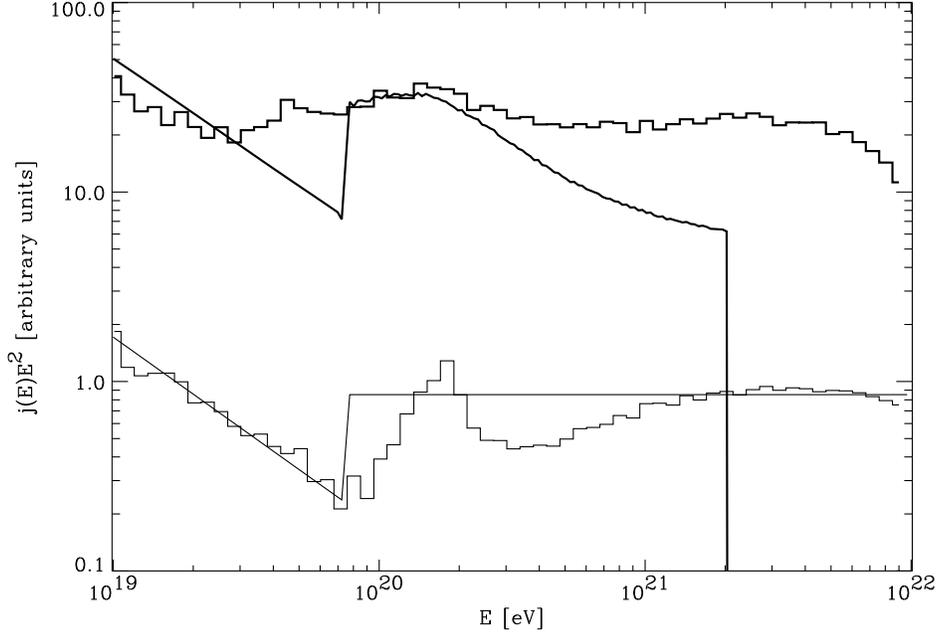}
\bigskip
\caption[...]{Same as Fig.~\ref{F1}, except that UHE CRs now propagate
in a sheet of finite half-thickness $5\,{\rm Mpc}$; both source and
observer are located close to the middle-plane.}
\label{F2}
\end{figure}

This main effect is shown as the thin solid histogram in Fig.~\ref{F2},
for a source distance $d=10\,{\rm Mpc}$, a sheet half-thickness
$d_s=5\,{\rm Mpc}$, an injection $\propto E^{-2}$, other physical
parameters as above, and no energy losses taken into account. The same
case with energy losses included is shown as the thick solid histogram.
The direct
comparison of this figure with Fig.~\ref{F1}, in the absence of energy
loss, shows that the flux of particles that diffuse in the sheet, is
depleted by  a factor $\simeq5$, with respect to the flux of
particles diffusing in an infinite medium. 

The shape of the spectrum
is, however, not significantly changed, and still goes as $Q(E)/D(E)$.
This is in agreement with the analytical solution for the sheet
geometry, assuming $j(E)=0$ at the sheet boundaries. For a point source
this solution which generalizes
Eq.~(\ref{Eq_diff_inf}) can be written as
\begin{equation}
j(E)=\frac{1}{4\pi}\sum_{m=0}^{+\infty}\,
\cos\left[\left(m+\frac{1}{2}\right)\,\pi\,\frac{z_o}{d_s}\right]
\cos\left[\left(m+\frac{1}{2}\right)\,\pi\,\frac{z_s}{d_s}\right]
g_m(r_o,r_s)\,,
\end{equation}
where $r_o$, $z_o$, $r_s$, $z_s$ respectively denote the cylindrical
coordinates of the observer and of the source, and $g_m(r_o,r_s)$ is
given, in the absence of energy losses, by
\begin{equation}
g_m(r_o,r_s)\,=\,\frac{Q(E)}{2\pi d_s D(E)}
{\rm K}_0\left[\left(m+\frac{1}{2}\right)\pi\frac{|r_o-r_s|}{d_s}\right]
\,,
\end{equation}
where ${\rm K}_0$ is a modified Bessel function of order 0, and, when
energy losses are included:
\begin{equation}
g_m(r_o,r_s,E)\,=\,\int_{0}^{+\infty}dE^\prime\,
\frac{Q(E^\prime)}{|b(E)|4\pi\lambda^2d_s}\exp\left[-
\left(m+\frac{1}{2}\right)^2
\pi^2\frac{\lambda(E,E^\prime)^2}{d_s^2}\right]
\exp\left[-\frac{(r_o-r_s)^2}{4\lambda(E,E^\prime)^2}\right]\,.
\end{equation}

These analytical solutions are shown in Fig.~\ref{F2} as light and
thick solid lines, respectively.
In this figure, the solution that takes energy
losses into account does not mix the two populations of particles,
$E>E_{\rm C}$ and $E<E_{\rm C}$, {\it i.e.}, ``secondary sources'' are not
included here, as they were for the infinite medium. However, as numerical
solutions of Eqs.~(\ref{cel_d}) and (\ref{subst}) show, inclusion of
secondary sources does not change the result significantly in this
case where $E_{\rm C}\lesssim70\,$EeV.
We note again that in our simulations, the nucleons are
propagated until either the distance traveled is larger than a Hubble
time, or twice the distance from the source to the observer. This means
that, in practice, 
the sheet is not infinite, and rather corresponds to a  disk,
of radius $2d$, and half-thickness $d_s$. It is possible to obtain
analytical solutions corresponding to that case, although such
expressions are rather cumbersome~\cite{B91}. Moreover, the above
solution should provide a relatively good approximation, as long as
$2d>d_s$, and we are here in this case. 

  Let us comment briefly on Fig.~\ref{F2}. In this figure, the
diffusion coefficient that gives the best fit between the analytical
solution and the Monte-Carlo is  a factor $\simeq8$ higher than the
Bohm diffusion coefficient,
to be compared with a factor $\simeq10$ in
the previous case (Fig.~\ref{F1}) of the infinite homogeneous medium.
This slight difference is due to the fact that, in the above analytical
solution, the diffusion coefficient is uniform in space, {\it i.e.} the
solution does not take into account the profile of the magnetic field
in the sheet. The diffusion coefficient, as averaged over the
thickness, is then $\simeq1.25$ times larger than the corresponding
uniform diffusion coefficient, for the same magnetic field strength
in the middle plane of the sheet.
In the infinite homogeneous case, there
is no such profile, and the above factor compensates for the difference
between the diffusion coefficients. We also note, in Fig.~\ref{F2},
that the transition between the diffusive and rectilinear regimes does
not appear to proceed smoothly. There is a peak around $\sim200\,$EeV,
which showed up for most spatial realizations of the magnetic field. We
discuss further below this dependence of the energy spectrum on the
particular realization of the turbulent magnetic field. The comparison
of the two cases, with and without energy losses, in Fig.~\ref{F2},
shows that this peak is smoothed out by energy losses, as expected.
Thus, when the geometry of the sheet is taken into
account, the above simple analytical solutions do not compare as well
to the Monte-Carlo simulations, as they did for the homogeneous case.

We wish to stress the importance of the particular geometry of the
sheet, as it precisely counteracts the effect of diffusion, in
depleting the flux at low energies. Several other effects,
which we now discuss, add up to this.

The profile of the magnetic field in the sheet is such that the field
is stronger in the middle plane than on the boundaries of the sheet.
Therefore, if source and observer are located on opposite sides of
the middle plane,
diffusive low energy particles are here as well depleted with respect
to non-diffusive high energy particles. This effect is not important
as long as source and observer are located within $\pm d_s/2$ of the
middle-plane, and, outside of this range, is dominated by the above
effect of free escape do the proximity to the boundary.
Therefore, we do not discuss it further.

\begin{figure}[ht]
\centering\leavevmode
\epsfxsize=5.0in
\epsfbox{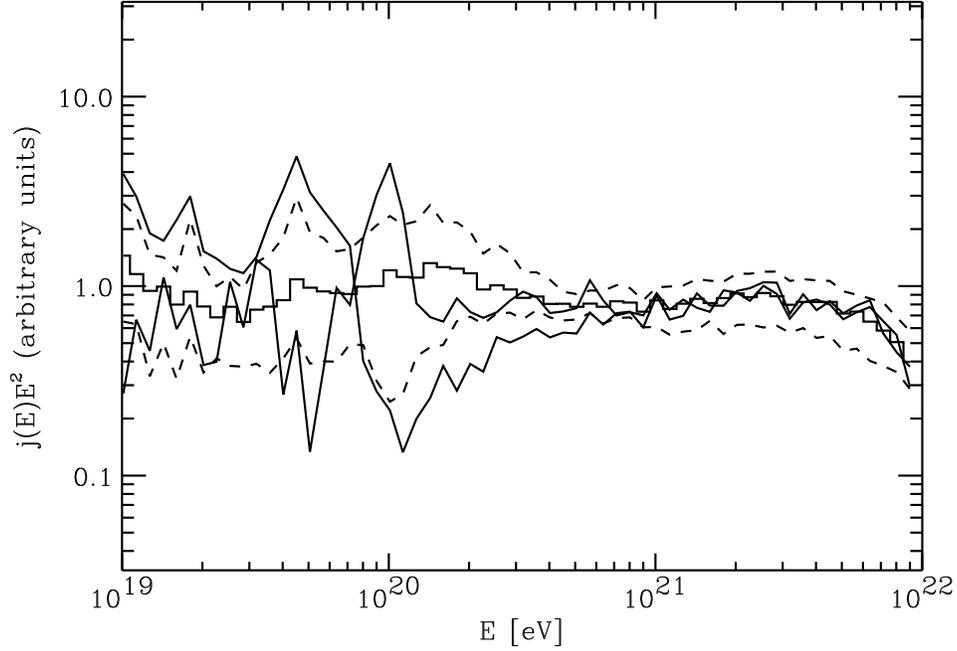}
\bigskip
\caption[...]{Spectrum of UHE CRs propagating in a sheet, with the same
parameters as in Fig.~\ref{F2}. The solid histogram shows the mean
flux, as obtained from the average of 13 fluxes corresponding to
different spatial realizations of the magnetic field; the dashed lines
show the upper and lower one-sided $1\sigma$ deviations from this mean.
The light
solid lines show two different spectra, corresponding to two different
realizations of the magnetic field.}
\label{F3}
\end{figure}

  In the presence of a sheet, the geometry of the problem becomes
highly anisotropic. It is all the more anisotropic when the distance to
the source is smaller or comparable to the largest turbulent eddy, 
{\it i.e}, $d\lesssim L\sim10\,{\rm Mpc}$ in our case, where most of the
power of the magnetic field lies. Indeed, particles of different
energies follow different paths, which, depending on the spatial
realization of the magnetic field, may, or may not, drive the particles
out of the sheet. As far as the UHE CR spectrum is concerned, this
effect translates itself in very substantial variations of the flux at
a given energy, in the diffusing regime, $\delta j(E)/j(E)\sim 1$. It
is  important to note that this effect does not represent an
overall variation of the flux, but is differential in energy. In
Fig.~\ref{F3}, we show two simulations each consisting of 4000 particles
in total, obtained from the same
parameters as above, that correspond to two different spatial
realizations of the magnetic field. These are shown in solid line. We
also show the mean flux, as obtained from the average of 13 such
simulations, {\it i.e.} 13 such realizations of the field, in solid
histogram. And, we also show the upper and lower $1\sigma$ deviations
from the mean, which show $\delta j(E)/j(E)\sim 1$. The two particular
realizations are not extreme in any respect, and certainly show that
the flux is highly dependent on the particular realization of the
magnetic field. This strong variation implies that analytical
solutions, which solve a diffusion equation that is already
spatially averaged over the realizations of the magnetic field,
are invalidated as long as they claim to solve a one
source -- one magnetic field model.
In the limit of high energies $E$, where the influence of the
magnetic field is negligible, the scatter $\delta j(E)$ should go
to zero. The finite
scatter around the mean flux visible in Fig.~\ref{F3} and the
subsequent ones is due to the finite numerical statistics of
4000 particles per field realization.

In the case of an infinite medium, there is also
substantial variation of the flux at a given energy; it is not as
strong, however, as when a sheet is present, for the reasons given
above. Our explanation is further supported by the fact that, for a
simulation where the power index of the magnetic field fluctuations is
positive, {\it i.e.}, where the power is concentrated on small
scales $\lambda_c\lesssim l_c=1$ Mpc, the relative deviation of the flux
from the mean is not significant. 

\begin{figure}[ht]
\centering\leavevmode
\epsfxsize=5.0in
\epsfbox{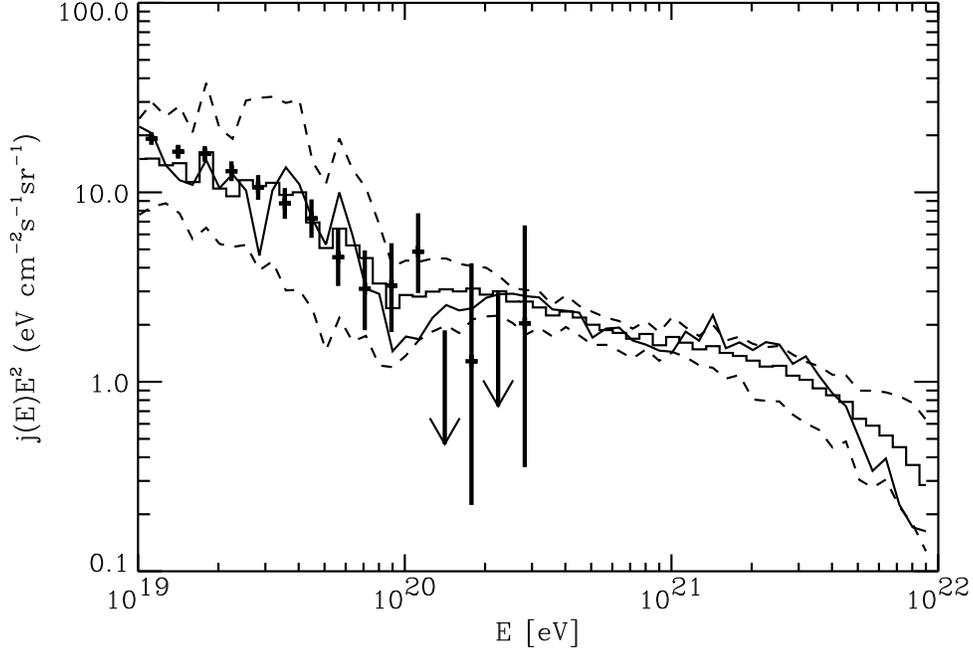}
\bigskip
\caption[...]{We show here the best fit obtained, for a magnetic field
$B_{\rm rms}\simeq5\times 10^{-8}\,{\rm G}$, a source distance
$d=5\,{\rm Mpc}$, and an injection spectrum $\propto E^{-2.4}$; the
data points are obtained from a compilation of the Haverah
Park~\cite{Haverah}, Fly's Eye~\cite{Bird1}, and AGASA~\cite{Yoshida1}
experiments. The solid histogram shows the mean flux,
obtained from the average of 16 different spatial realizations of the
turbulent magnetic field, while the light solid line corresponds to
one, carefully chosen, best fit, realization of the field.}
\label{F4}
\end{figure}

Qualitatively speaking, if one wishes to reproduce the observed UHE
CR spectrum with the help of diffusion in the Local Supercluster, there
are several paths to follow. In either case, one has to find a way to
increase the flux of UHE CR at the lowest energies 
$\lesssim 50\,{\rm EeV}$. Therefore, one could select a situation in
which $d<d_s$, so that the boundary of the sheet does not play a
significant role. The closest energetic galaxy is Centaurus A, at a
distance $\simeq5\,{\rm Mpc}$. In Fig.~\ref{F4}, we show the best fit we
were able to obtain for $d=5\,{\rm Mpc}$, out of 16 spatial
realizations of the magnetic
field, using an injection spectrum $\propto E^{-2.4}$ in order to
match the observations; other physical parameters are as above. All
other spatial realizations did not give a satisfying agreement with
observations, and, even in this case, the fit is only barely acceptable.
However, we certainly cannot exclude all spatial realizations as
consistent solutions, in the present case. Our simulations only show
that such a solution, if it exists, is {\it a priori} unlikely, where
the {\it a priori} refers to the statistical distribution of field
realizations.

\begin{figure}[ht]
\centering\leavevmode
\epsfxsize=5.0in
\epsfbox{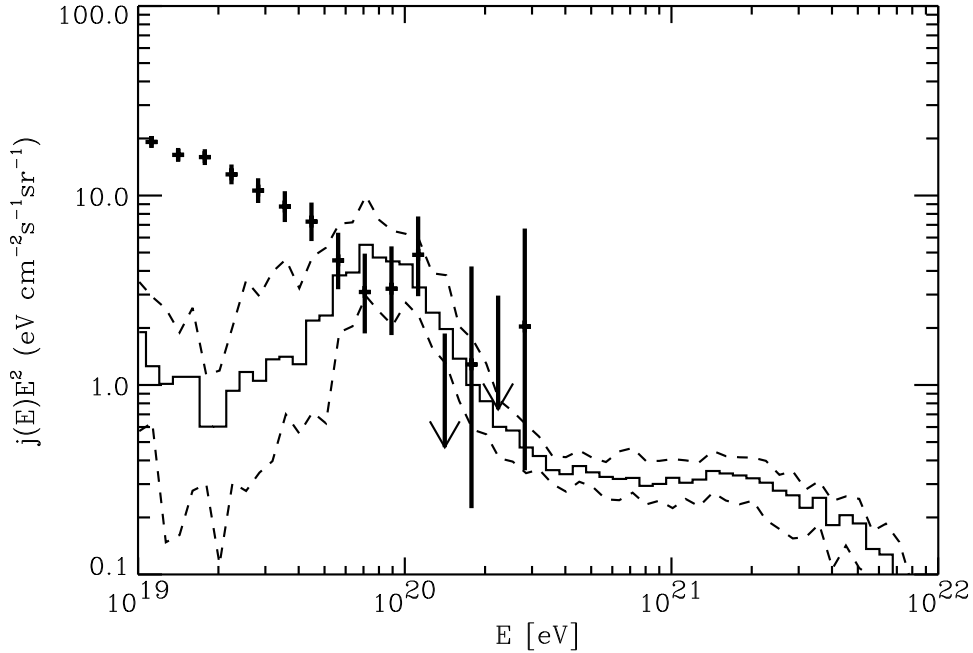}
\bigskip
\caption[...]{Spectrum of UHE CRs propagating in the Local
Supercluster, for a source distance $d=25\,{\rm Mpc}$, other physical
parameters as in Fig.~\ref{F4}. The solid histogram shows the mean
flux, as obtained from the average of fluxes resulting in 11
different spatial realizations of the magnetic field; the dashed lines
show the upper and lower one-sided $1\sigma$ deviations from this mean.}
\label{F5}
\end{figure}

Another way to follow would be, loosely speaking, to increase the
energy losses so as to increase the flux at low energies with respect
to that at high energies. One should not increase the distance,
however, as  free escape would then affect the low
energy part of the spectrum
even more dramatically. This is shown in Fig.~\ref{F5}, where the flux
is shown for a source distance $d=25\,{\rm Mpc}$, all other parameters
as above. However, if one increases the strength of the magnetic field,
particles with energies higher than the GZK cut-off start to diffuse.
Diffusion increases the distance traveled to $d + \tau_E$.
Pion production is then affected by the magnetic field as soon
as $\tau_E\sim d$, which occurs, for $E\gtrsim50\,$EeV, for 
$B_{\rm rms}\gtrsim 10^{-7}\,$G, and $d\sim10\,$Mpc,
as seen from Eq.~(\ref{delayrect}).
Furthermore, a stronger magnetic field results in a smaller diffusion
coefficient, which, in turn, implies a higher number density of
diffusing particles, all things being equal. Let us now furnish this
discussion with more quantitative arguments.

\begin{figure}[ht]
\centering\leavevmode
\epsfxsize=5.0in
\epsfbox{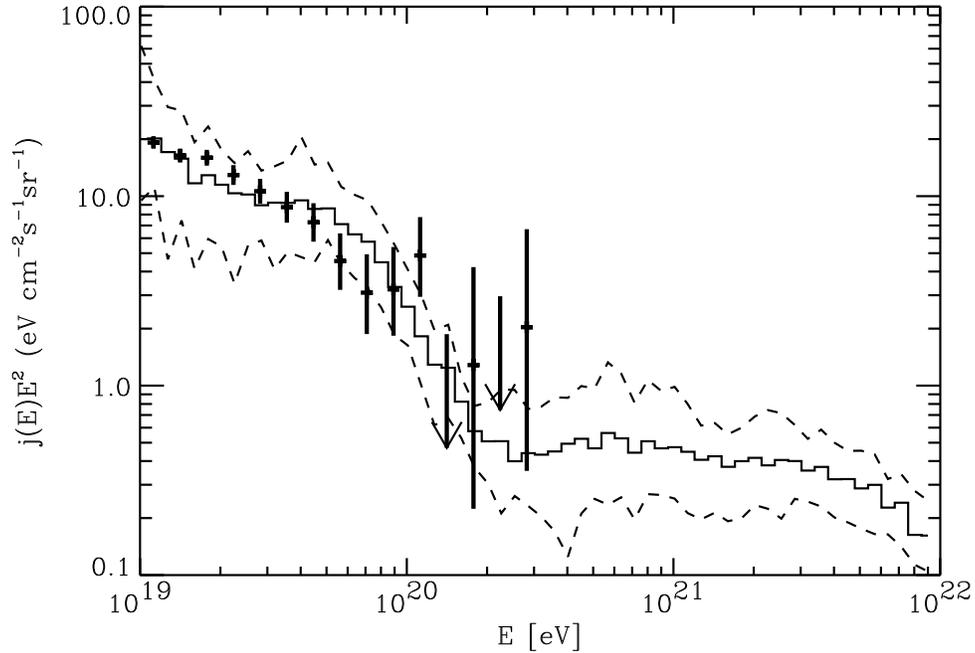}
\bigskip
\caption[...]{Spectrum of UHE CRs, as in Fig.~\ref{F5}, but for a
source distance $d=10\,{\rm Mpc}$, and a magnetic field
$B_{\rm rms}\simeq 3\times10^{-7}\,{\rm G}$, averaged over 10
field configurations.}
\label{F6}
\end{figure}

\begin{figure}[ht]
\centering\leavevmode
\epsfxsize=5.0in
\epsfbox{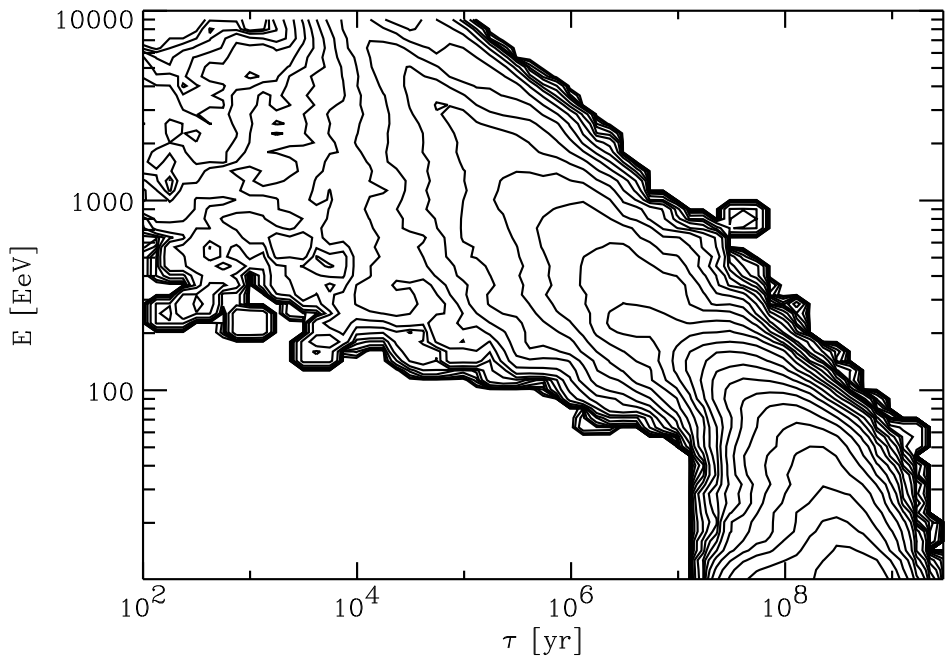}
\bigskip
\caption[...]{The distribution of time delays $\tau_E$ and energies
$E$ for the case shown in Fig.~\ref{F6}.
The inter-contour interval is 0.25 in the logarithm to base 10 of the
distribution per logarithmic energy and time interval.
The three regimes discussed in the text,
$\tau_E\propto E^{-2}$ in the rectilinear regime $E\gtrsim200\,$EeV,
$\tau_E\propto E^{-1}$ in the Bohm diffusion regime $60\,{\rm EeV}
\lesssim E\lesssim200\,$EeV, and $\tau_E\propto E^{-1/3}$ for
$E\lesssim60\,$EeV are clearly visible.}
\label{F7}
\end{figure}

In Fig.~\ref{F6}, we show an extreme case, where
$B_{\rm rms}\simeq3\times10^{-7}\,{\rm G}$, and the injection
spectrum $\propto E^{-2.4}$. There, particles up to
$\sim200\,{\rm EeV}$ diffuse, and the effective distance that they
travel is $\sim 1.4\times d\sim14\,{\rm Mpc}$ for
$E\sim 100\,{\rm EeV}$, and $\sim 3\times d$ for $E\sim 70\,{\rm EeV}$,
with a large relative scatter in the distance traveled of order 1, at
all energies. In this extreme case, pion production is very strong:
particles are strongly depleted at high energies due to energy
losses, while lower energy particle abundances
are increased due to diffusion and energy losses of higher
energy particles. Because all three regimes in the correlation
between $\tau_E$ and $E$ described in the previous section
appear in this case above $10\,$EeV, we show in Fig.~\ref{F7},
the corresponding distribution of UHE CRs in the time
delay--energy plane. There one can see the correlation
$\tau_E\propto E^{-2}$ at high energies $E\gtrsim 200\,{\rm EeV}$,
and the associated scatter, that results from stochastic pion
production losses. Below these energies, where particles diffuse,
the correlation switches to $\tau_E\propto E^{-1}$ in the Bohm regime,
and to $\tau_E\propto E^{-1/3}$ for $E\lesssim60\,$EeV. Note that
the gyro radius of a $60\,{\rm EeV}$ proton in a
$3\times10^{-7}\,{\rm G}$ magnetic field is
$r_g\simeq200\,{\rm kpc}\simeq0.1 L/(2\pi)$. Our simulations thus
indicate that the $D(E)\propto E^{1/3}$ correlation only arises when 
$r_g<0.1L/(2\pi)$, and $D(E)\propto E$ otherwise. Furthermore,
Figs.~\ref{F6} and~\ref{F7} indicate that the transition regime
between diffusion and rectilinear propagation is rather
narrow, a factor 2--3 in energy. This is consistent with the
two estimates of the transition energy $E_{\rm C}$ from
Eq.~(\ref{Ecrit}) and from $d=D(E_{\rm C})$, {\it if} the diffusion
coefficient derived from the simulations,
$D(E)\simeq10D_{\rm Bohm}(E)$, is used in the latter.

\begin{figure}[ht]
\centering\leavevmode
\epsfxsize=5.0in
\epsfbox{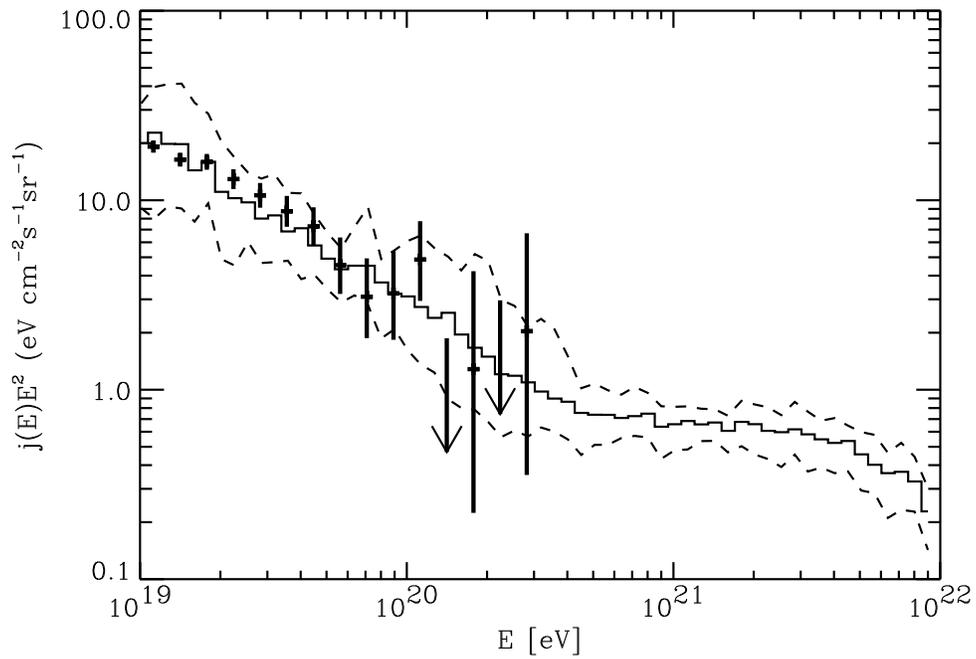}
\bigskip
\caption[...]{Final best fit solution, obtained for a source distance
$d=10\,{\rm Mpc}$, and a magnetic field strength 
$B_{\rm rms}\simeq 10^{-7}\,{\rm G}$, other physical parameters and
line key as in Fig.~\ref{F5}, where the histogram was averaged over
15 field configurations.}
\label{F8}
\end{figure}

The best, and consistent, fit to the data is obtained for the
intermediate value
of the magnetic field strength, $B_{\rm rms}\simeq10^{-7}\,{\rm G}$,
and an injection spectrum $\propto E^{-2.4}$. Amazingly, this injection
spectrum has been derived for UHE CRs from clusters of
galaxies~\cite{ensslin2}. In Fig.~\ref{F8}, we show
the resulting UHE CR spectrum, with the associated upper and lower
$1\sigma$ deviations. Most realizations of the magnetic field give, in
this case, acceptable fits to the observed spectrum.

\begin{figure}[ht]
\centering\leavevmode
\epsfxsize=7in
\epsfbox{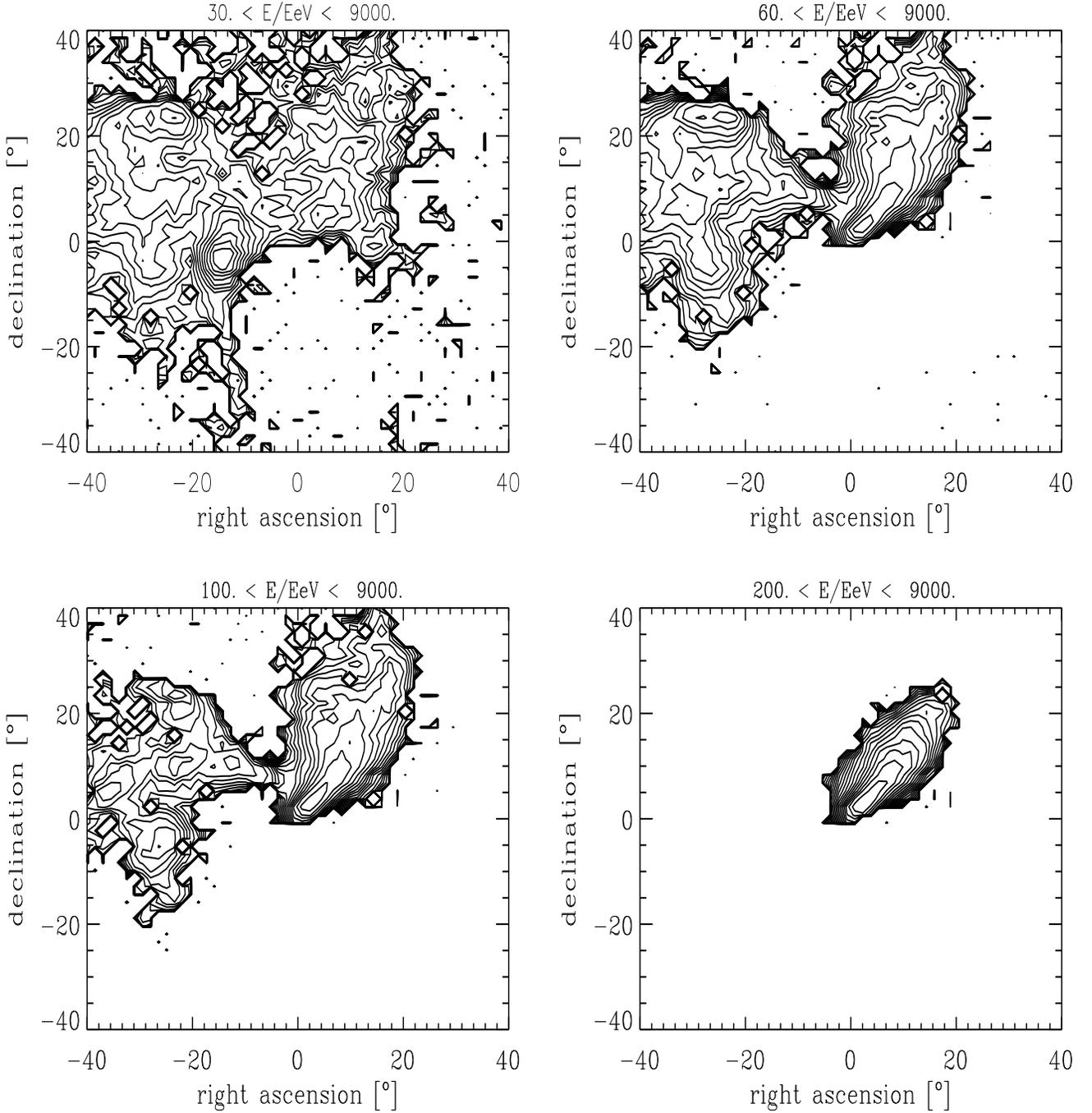}
\bigskip
\caption[...]{Angular image of a point-like source, corresponding to
one particular realization of the scenario shown in Figs.~\ref{F2}
and \ref{F3}, as seen by a detector of $\simeq1^\circ$ angular
resolution, in different energy
ranges, as indicated. An $E^{-2.4}$ injection spectrum was assumed.
A transition
from several images at lower energies to only one image at the
highest energies occurs where the linear deflection becomes
comparable to the effective field coherence length, as discussed in the
text. The inter-contour interval is 0.1 in the logarithm to base 10
of the integral flux per solid angle.}
\label{F9}
\end{figure}

For the sake of illustration, we show in Fig.~\ref{F9}, the
angular images at different energies, as indicated, associated with
a typical realization corresponding to the case shown in Figs.~\ref{F2}
and~\ref{F3}, but for our standard injection spectrum with $\gamma=2.4$.
Interestingly, a cross-over from several images of the same source
at low energies to only one image at the highest energies
can be observed. This transition occurs at an energy for which
the linear deflection $d\theta_E$ ($\theta_E$ being the deflection
angle) becomes comparable to the field coherence scale $\lambda_c$.
This would in turn allow to estimate $\lambda_c$.
Furthermore, for $E\gtrsim 200\,{\rm EeV}$, the typical deflection
angle is $\simeq13^\circ$.
This gives the order of magnitude of the angular box in
which one should look for counterparts associated with the highest
energy events observed.

Finally, for continuously emitting sources, the source power $Q$ required to
reproduce the spectra discussed here is easily estimated as
\begin{eqnarray}
  Q\equiv\int dE Q(E)=&\sim&(4\pi)^2d^2E^2j(E)\max
  \left[\left(E_{\rm max}/E\right)^{(2-\gamma)},
  \left(E_{\rm min}/E\right)^{(2-\gamma)}\right]
  \nonumber\\
  &\sim&
  7.2\times10^{45}\left(\frac{d}{10\,{\rm Mpc}}\right)^2
  \max\left[\left(E_{\rm max}/E\right)^{(2-\gamma)},
  \left(E_{\rm min}/E\right)^{(2-\gamma)}\right]
  \,{\rm erg}\,{\rm s}^{-1}\label{power}
\end{eqnarray}
at energies $E$ where deflection is small.
Here, $E_{\rm min}$ is the low energy cut-off of the injection
spectrum (relevant if $\gamma>2$), and the second expression
has been obtained from the observed flux at $E\simeq300\,$EeV. This
model, for a single source, active over $\gtrsim10^8\,$yr, thus requires
a very high cosmic ray luminosity, in agreement with previous
estimates~\cite{WW79}. For comparison, the observed luminosity of NGC
4151, a powerful Seyfert galaxy of the local Supercluster, located at a
distance $\sim13\,$Mpc (for $H_0=75\,$km/s/Mpc), is
$L\sim10^{45}\,$erg/s, and the typical bolometric luminosity of a
Seyfert galaxy is $L\sim10^{44}-10^{46}\,$ erg/s. On the theoretical
side, expectations for the UHE CR luminosity are uncertain and
vary greatly with the type of
accelerator and the emission timescale, {\it e.g.},
$L_{\rm UHECR}\sim10^{46}\,$erg/s
for acceleration in Active Galactic Nuclei over $\sim10^8\,$yr, 
$L_{\rm UHECR}\sim10^{51}\,$erg/s over $\sim100\,$s for fireball
models of $\gamma-$ray bursts, up to $L_{\rm UHECR}\gtrsim10^{52}\,$erg/s
over $\gtrsim1\,{\rm s}-10^4\,{\rm yr}$ (spread in photons, neutrinos,
and nucleons), for topological defects, and up to $L_{\rm UHECR}\sim
3\times10^{47}\,$erg/s for radiogalaxies~\cite{falcke}.

  A one source model, such as considered above, certainly demands a huge
output of energy in UHE CRs, when one integrates the required luminosity
over the required emission timescale. Nevertheless, we note that
intermittent sources, with pulses of duration $T\ll\tau_E\sim10^8\,$yr, with
inter-pulse durations $\Delta T\ll\tau_E$, and, preferably $T\ll\Delta T$,
would mimic a continuously emitting source, as the scatter in time delays
$\delta\tau_E\sim\tau_E\gg\Delta T$. Obviously, the energy budget would
then be greatly alleviated. In fact, the hot spots in radiogalaxies
that are believed to be UHE CR sources~\cite{biermann2} show
intermittent behavior on time scales of $\sim10^4\,$yr and smaller.
Also, nothing prevents us from
considering more than one source; for instance, there are at least seven
powerful Seyfert galaxies in the Local Supercluster, whose
distance lies between $\sim5\,$Mpc and $\sim20\,$Mpc. As well,
the combination of many
bursting sources, with emission timescales $\ll10^{8}\,$yr, would
reproduce a spectrum similar to that of a single continuously emitting
source.

  Finally, we note that the injection spectrum index that we
considered here, $\gamma=2.4$, is very close to the value expected
from acceleration in ultra-relativistic shocks
$\simeq2.2\pm0.1$~\cite{Ostrowski98}.

\section{Conclusions}

We have discussed the propagation of charged UHE CR nucleons in a
locally enhanced extra-galactic magnetic field of strength 
$B_{\rm rms}\gtrsim 10^{-8}\,{\rm G}$, in the Local Supercluster. Such
high values for the ambient magnetic field in walls and filaments of
large scale structure obey observational constraints such as
Faraday rotation measurements and seem to be not unrealistic,
as recently advertised by Ryu, Kang \&
Biermann~\cite{RKB98}. In this regime, UHE CRs diffuse up to 
$\sim 100\,{\rm EeV}$ for $B_{\rm rms}\sim 10^{-7}\,{\rm G}$, and a
distance to the source $d\sim10\,{\rm Mpc}$, and
propagate recti-linearly beyond, as first
advocated by Wdowczyk \& Wolfendale~\cite{WW79}. As a consequence,
for sources emitting on timescales large  compared to typical
propagation times, the spectrum in the diffusive range is steeper than
the injection spectrum, and typically scales as 
$Q(E) E^{-0.6}$, with $Q(E)$ the injection spectrum.
This is a result of an interplay between the energy dependent
diffusion coefficient and energy loss rates. For sources emitting
on smaller timescales, the spectrum tends to be peaked at
a certain energy that depends on the turn-on time of the
source. Diffusion models present
many advantages in the context of UHE CRs. For instance, the switch off
of diffusion at high energy might provide a simple way to reconcile
the seemingly observed flattening of the UHE CR spectrum above the GZK
cut-off with a single power law injection spectrum
that is relatively soft, $Q(E)\propto E^{-2.4}$. Moreover,
the large deflection angles associated with the strong magnetic fields
could explain why no astrophysical counterpart could be associated with
the highest energy event, within a $\simeq2^\circ$ error box.

The problem has been recently reconsidered in Ref.~\cite{BO}, using
an analytical solution of the diffusion equation in an infinite medium,
matched to the near straight line propagation regime at high energy.
Here, we have performed numerical simulations of the full problem,
including a plane geometry model of the structure of the local
Supercluster, and stochastic pion production
energy losses. Our simulations improve on analytical solutions, as they
allow to simulate the transition between
the diffusive and the non-diffusive regimes, and
the non-trivial geometry (here modeled as a planar sheet), with
inhomogeneity length scales for the density and magnetic fields of,
typically, a few Mpc.

Our results differ from the analytical solutions of diffusion in an
infinite medium. As a main effect, we find
that the free escape of particles at the boundary,
due to the negligible magnetic field
outside of the sheet, depletes the low energy diffusive part of the UHE
CR spectrum with respect to the high energy (non-diffusive) part.
For a distance
comparable to the thickness of the sheet, this depletion is  a
factor $\simeq5$ as compared to the flux predicted for particles
diffusing in
an infinite medium (no boundary). We also find that the highly
anisotropic geometry of the problem, which is all the more
anisotropic when the source
distance is smaller or comparable to the largest turbulent eddy,
implies a strong relative variation in flux at any energy in the
diffusing part, of order 1. Furthermore, our simulations recover
the scaling of the diffusion coefficient indicated by analytical
estimates, but with about a factor 10 higher normalization both
in the infinite medium and for the sheet geometry.

We thus find that, for sources emitting over timescales larger than
$\sim 10^8\,{\rm yr}$: {\it (i)} if the magnetic field is not strong
enough, {\it i.e.} $B_{\rm rms}\lesssim 5\times10^{-8}\,{\rm G}$, the
diffusion coefficient is not small enough to compensate the depletion
of low energy diffusive particles by ejection out of the sheet; no
entirely satisfying fit to the observed UHE CR spectrum can be obtained
in this case. This result holds even for small distances,
down to $d\sim 5\,{\rm Mpc}$, as the sheet half-thickness is only
$\sim5\,{\rm Mpc}$; {\it (ii)} if the source distance is
significantly larger than the thickness of the sheet
$\sim10\,{\rm Mpc}$, or if source and/or
observer are located near the boundary of the sheet, within 
$\sim\pm 2\,{\rm Mpc}$, there is no consistent fit to the observed UHE
CR spectrum. Indeed, in this case, pion production becomes significant
and depletes the high energy part of the spectrum, {\it i.e.} above the
GZK cut-off, while the diffusive particles feel quite strongly the
presence of the sheet boundary. 

  However, we find a very nice and consistent fit to the observed UHE
CR spectrum for a source distance $d\sim10\,{\rm Mpc}$, and a magnetic
field $B_{\rm rms}\sim10^{-7}\,{\rm G}$. Smaller magnetic field
strengths suffer from the above problem, while stronger magnetic field
imply too much pion production at high energies. Indeed, for the above
magnetic field, particles up to $\sim 100\,{\rm EeV}$ diffuse, and
diffusion considerably increases the effective distance traveled. We
thus obtain a solution only in a limited region of parameter space.

  A magnetic field $B_{\rm rms}\sim10^{-7}\,{\rm G}$ may not be a
coincidence after all, as the equipartition field expected in walls
of large scale structure, such as the Local Supercluster, is
$B_{\rm equi}\sim 0.3\,\mu{\rm G}\,h_{100}
\left(T/3\times10^6\,{\rm K}\right)^{1/2}
\left(\rho_{\rm b}/0.3\rho_{\rm c}\right)^{1/2}$~\cite{Kronberg,RKB98}, where
$T$ and $\rho_{\rm b}$ denote respectively the temperature and baryon
density within the sheet, while $\rho_{\rm c}$ is the critical density,
and $h_{100}$ the value of the Hubble constant in units of 100~km/s/Mpc.

Based on the technique developed here, we will perform detailed
studies of spectra and angular images of ultra-high energy
cosmic ray sources for more realistic source and magnetic field
distributions in subsequent work.

Next generation experiments with their potential to record tens to
hundreds of events above $100\,$EeV will be able to detect these
signatures and exploit the information associated with them.

\section*{Acknowledgments}
We especially thank the late David Schramm for constant encouragement
and collaboration in earlier work. Pasquale Blasi, Chris Hill,
Angela Olinto, and J\"org Rachen are acknowledged for valuable discussions. 
We are grateful to the Max-Planck Institut f\"ur Physik, M\"unchen
(Germany), and the Institut d'Astrophysique de Paris, Paris (France),
for providing CPU time. This work was supported, in
part, by the DoE, NSF, and NASA at the University of Chicago.

\newpage

\end{document}